\documentclass[aps,amsmath,amssymb,reprint,twocolumn,pra,superscriptaddress,notitlepage,showpacs,tightenlines]{revtex4-1}

\usepackage{graphicx}% Include figure files
\usepackage{dcolumn}% Align table columns on decimal point
\usepackage{bm}% bold math
\usepackage{textcomp}
\usepackage[utf8]{inputenc}
\usepackage[T1]{fontenc}
\usepackage{mathptmx}
\usepackage{appendix}
\usepackage{color}
\usepackage{textcomp}
\usepackage{amsmath}
\usepackage{float}
\usepackage[colorlinks,citecolor=blue,linkcolor=red,hyperindex,CJKbookmarks]{hyperref}

\allowdisplaybreaks[1]

\begin{document}

\title{Enantio-detection of cyclic three-level chiral molecules in a driven cavity}

\author{Yu-Yuan Chen}
\affiliation{Beijing Computational Science Research Center, Beijing 100193, China}

\author{Jian-Jian Cheng}
\affiliation{Beijing Computational Science Research Center, Beijing 100193, China}

\author{Chong Ye}
\affiliation{Beijing Key Laboratory of Nanophotonics and Ultrafine Optoelectronic Systems, School of Physics, Beijing Institute of Technology, 100081 Beijing, China}

\author{Yong Li}
\email{liyong@csrc.ac.cn}
\affiliation{Beijing Computational Science Research Center, Beijing 100193, China}

\date{\today}

\begin{abstract}
We propose an enantio-detection method of chiral molecules in a cavity with external drive. The chiral molecules are coupled with a quantized cavity field and two classical light fields to form the cyclic three-level systems. The chirality-dependent cavity-assisted three-photon process in the three-level systems leads to the generation of intracavity photons. Simultaneously, the drive field also results in the chirality-independent process of the generation of intracavity photons. Based on the interference between the intracavity photons generated from these two processes, one can detect the enantiomeric excess of chiral mixture via monitoring the transmission rate of the drive field.
\end{abstract}
\maketitle

\section{Introduction}
The existence of two molecular structural forms known as enantiomers (left- and right- handed chiral molecules) is one of the most important manifestations of symmetry breaking in nature~\cite{MolecularStructure}. Chiral molecules refer to the molecules that cannot be superposed on their mirror images via translations and rotations. They play crucial roles in various enantio-selective biological activities and chemical reactions~\cite{EnantioEffect-Chiralitye2012,EnantioEffect-Science1996}. Thus, enantio-detection~\cite{Barron-MolecularScatter,Discrimination-JSepSci2007,Discrimination-OR,Discrimination-VCD,Discrimination-JiaWZ,Discrimination-ZhangXD,Discrimination-Lehmann,Discrimination-YeC,Discrimination-XuXW,Discrimination-ChenYY,Discrimination-Lehmann-Arxiv1,Microwave-Doyle-Nature,Microwave-Doyle-PRL,Microwave-Lehmann-JPCL,Microwave-Schnell-ACIE,Microwave-Schnell-JPCL,Discrimination-ZhangX}~of chiral molecules is an important and challenging work. Most conventional spectroscopic methods~\cite{Barron-MolecularScatter,Discrimination-JSepSci2007,Discrimination-OR,Discrimination-VCD}~for enantio-detection of chiral molecules are based on the interference between the electric-dipole and magnetic-dipole (or electric-quadrupole) transitions, and thus, the chiral signals are weak since the magnetic-dipole and electric-quadrupole transition moments are usually weak compared with the electric-dipole transition moments.

Recently, the cyclic three-level systems~\cite{Discrimination-JiaWZ,Discrimination-Lehmann,Discrimination-YeC,Discrimination-XuXW,Discrimination-ChenYY,Microwave-Doyle-Nature,Microwave-Doyle-PRL,Microwave-Lehmann-JPCL,Microwave-Schnell-ACIE,Microwave-Schnell-JPCL,Discrimination-Lehmann-Arxiv1,ThreeLevel-LiuYX,ThreeLevel-LiY,ThreeLevel-ZhouL,ThreeLevel-Hirota,ThreeLevel-YeC,ThreeLevel-Vitanov,ThreeLevel-WuJL1,Spatial-Separation-LiY-PRL,Spatial-Seperation-Hornberger-JCP,Spatial-Seperation-Shapiro-JCP,Spatial-Seperation-LiuB,Seperation-Shapiro-PRL,Seperation-JiaWZ-JPB,Seperation-Koch-JCP,Seperation-LiY-PRA,Seperation-Schnell-ACIE,Seperation-SepDoyle-PRL,Seperation-Vitanov-PRR,Seperation-YeC-PRA,Seperation-ZhangQS-JPB,Spatial-Seperation-Kravets-PRL,Spatial-Seperation-Cipparrone-LSAppl,Discrimination-ZhangX}~of chiral molecules involving only the electric-dipole transitions have been widely used in enantio-detection~\cite{Discrimination-JiaWZ,Discrimination-Lehmann,Discrimination-YeC,Discrimination-XuXW,Discrimination-ChenYY,Microwave-Doyle-Nature,Microwave-Doyle-PRL,Microwave-Lehmann-JPCL,Microwave-Schnell-ACIE,Microwave-Schnell-JPCL,Discrimination-Lehmann-Arxiv1,Discrimination-ZhangX}, enantio-specific state transfer~\cite{Seperation-Shapiro-PRL,Seperation-JiaWZ-JPB,Seperation-Koch-JCP,Seperation-LiY-PRA,Seperation-Schnell-ACIE,Seperation-SepDoyle-PRL,Seperation-Vitanov-PRR,Seperation-YeC-PRA,Seperation-ZhangQS-JPB}, and spatial enantio-separation~\cite{Spatial-Separation-LiY-PRL,Spatial-Seperation-Hornberger-JCP,Spatial-Seperation-Shapiro-JCP,Spatial-Seperation-LiuB,Spatial-Seperation-Kravets-PRL,Spatial-Seperation-Cipparrone-LSAppl}~of chiral molecules. Specially, the enantiomer-specific microwave spectroscopic methods~\cite{Microwave-Doyle-Nature,Microwave-Doyle-PRL,Microwave-Lehmann-JPCL,Microwave-Schnell-ACIE,Microwave-Schnell-JPCL}~based on the cyclic three-level systems have achieved great success in the investigations of enantio-detection of chiral molecules. Due to the inherent properties of electric-dipole transition moments of enantiomers, the product of three electric-dipole transition moments for the cyclic three-level systems changes sign with enantiomer. Thus, when the molecules in chiral mixture are coupled with two classical light fields, the total induced light field generated via the three-photon process of three-wave mixing~\cite{Microwave-Doyle-Nature,Microwave-Doyle-PRL,Microwave-Lehmann-JPCL,Microwave-Schnell-ACIE,Microwave-Schnell-JPCL}~is determined by the difference between the numbers of left- and right- handed molecules. Consequently, one can detect the enantiomeric excess of the chiral mixture via monitoring the intensity of the total induced light field.

On the other hand, cavity quantum electrodynamics (CQED) systems with a single molecule or many molecules confined in a cavity have received considerable interest~\cite{CQEDMolecule2,CQEDtransfer1,CQEDtransfer2,CQEDtransfer3,CQEDMoleculeSpectra1,CQEDMoleculeSpectra2,CQEDMoleculeSpectra3,CQEDReaction1,CQEDReaction2,CQEDReaction3,CQEDMoleculeXia,CQEDMoleculeWe}. In such CQED systems, the electromagnetic environment of the molecule(s) is modified by the quantized cavity field. This can strengthen the interaction between light fields and molecule(s) dramatically. Thus, the CQED systems have shown promising applications in the fields of energy transfer~\cite{CQEDtransfer1,CQEDtransfer2,CQEDtransfer3}, molecular spectra~\cite{CQEDMoleculeSpectra1,CQEDMoleculeSpectra2,CQEDMoleculeSpectra3}, and control of chemical reactions~\cite{CQEDReaction1,CQEDReaction2,CQEDReaction3}~for molecules.

Most recently, the enantio-detection of single chiral molecule~\cite{CQEDMoleculeXia}~or many chiral molecules~\cite{CQEDMoleculeWe}~has been investigated theoretically in the CQED systems. In Ref.~\cite{CQEDMoleculeXia}, it has been proposed to distinguish the chirality of single chiral molecule by using the single-molecule model of cyclic three-level system. However, in realistic case, the systems of enantio-detection~\cite{Barron-MolecularScatter,Discrimination-JSepSci2007,Discrimination-OR,Discrimination-VCD,Discrimination-JiaWZ,Discrimination-ZhangXD,Discrimination-Lehmann,Discrimination-YeC,Discrimination-XuXW,Discrimination-ChenYY,Discrimination-Lehmann-Arxiv1,Microwave-Doyle-Nature,Microwave-Doyle-PRL,Microwave-Lehmann-JPCL,Microwave-Schnell-ACIE,Microwave-Schnell-JPCL}~(as well as enantio-specific state transfer~\cite{Seperation-Shapiro-PRL,Seperation-JiaWZ-JPB,Seperation-Koch-JCP,Seperation-LiY-PRA,Seperation-Schnell-ACIE,Seperation-SepDoyle-PRL,Seperation-Vitanov-PRR,Seperation-YeC-PRA,Seperation-ZhangQS-JPB}, spatial enantio-separation~\cite{Spatial-Separation-LiY-PRL,Spatial-Seperation-Hornberger-JCP,Spatial-Seperation-Shapiro-JCP,Spatial-Seperation-LiuB,Spatial-Seperation-Kravets-PRL,Spatial-Seperation-Cipparrone-LSAppl}, and enantio-conversion~\cite{Conversion-Cohen-PRL,Conversion-Shapiro-JCP,Conversion-Shapiro-PRL,Conversion-YeC-PRR,Conversion-YeC1,Conversion-YeC2,Conversion-Shapiro-PRA}) of chiral molecules commonly contain a large number of molecules. In the case of the quantized cavity field(s) coupling with many molecules, one should resort to the multi-molecules treatment~\cite{CQEDtransfer1,CQEDtransfer2,CQEDMoleculeSpectra1,CQEDMoleculeSpectra2,CQEDMoleculeSpectra3,CQEDReaction1,CQEDReaction2,CQEDMoleculeSpectra2,CQEDReaction3,CQEDtransfer3}, rather than the single-molecule one, which is only appropriate in the case of classical field(s) interacting with many molecules~\cite{Barron-MolecularScatter,Discrimination-JSepSci2007,Discrimination-OR,Discrimination-VCD,Discrimination-JiaWZ,Discrimination-ZhangXD,Discrimination-Lehmann,Discrimination-YeC,Discrimination-XuXW,Discrimination-ChenYY,Discrimination-Lehmann-Arxiv1,Microwave-Doyle-Nature,Microwave-Doyle-PRL,Microwave-Lehmann-JPCL,Microwave-Schnell-ACIE,Microwave-Schnell-JPCL}. In Ref.~\cite{CQEDMoleculeWe}, the enantio-detection of chiral mixture has been achieved with the multi-molecule treatment in the CQED system for cyclic three-level chiral molecules in a cavity without external drive. The chirality-dependent cavity-assisted three-photon process leads to the generation of intracavity photons (even in the absence of external drive to the cavity). Thus it provided a promising way to detect the enantiomeric excess of chiral mixture by measuring the output field of the cavity.

In this paper, we propose an enantio-detection method based on the CQED system for cyclic three-level chiral molecules, which locate in a traveling-wave cavity~\cite{RingCavity-Xiao-2001,RingCavity-Xiao-2008,RingCavity-Culver-2016}~with external drive. Each molecule is described by the cyclic three-level system coupled with the quantized cavity field and two classical light fields. In the absence of the external drive, due to the existence of the two classical light fields, the intracavity photons can be generated via the chirality-dependent cavity-assisted three-photon process~\cite{CQEDMoleculeWe}. In the presence of the external drive, the drive field enters the cavity and also results in the chirality-independent process of generation of intracavity photons. There exists the interference between the intracavity photons resulting from these two processes. Based on this, we demonstrate that the enantiomeric excess can be detected by monitoring the steady-state transmission rate of the drive field.

We remark that in the previous system~\cite{CQEDMoleculeWe}~where the cyclic three-level model is designed in the standing-wave cavity, the size of sample is required to be much smaller than the wavelengths of all the light fields to evade the influence of the phase-mismatching and the spatial dependence of the coupling strength. In our current system, however, such a strict requirement is not necessary since the present cyclic three-level model is specially-designed in the traveling-wave cavity. On the other hand, there is no external drive to the cavity in the previous system~\cite{CQEDMoleculeWe}~and thus the enantiopure samples are required in the enantio-detection therein. In contrast, the existence of the external drive to the cavity in our system ensures our present method works without requiring the enantiopure samples. Therefore, the present method has advantages in enantio-detection of chiral molecules compared with the previous one in Ref.~\cite{CQEDMoleculeWe}.

This paper is organized as follows. In Sec.~\ref{ModEq}, we give the model and Hamiltonian of the CQED system for cyclic three-level chiral molecules. Then the steady-state transmission of the drive field is investigated in Sec.~\ref{transmissivity}. Further, we present the results for enantio-detection of chiral molecules in Sec.~\ref{DetectEE}, and then give the discussions about our investigations in Sec.~\ref{discussion}. Finally, the conclusion is given in Sec.~\ref{summary}.

\section{Model and Hamiltonian}\label{ModEq}
\begin{figure}[tbp]
	\centering
	\includegraphics[width=8.5cm]{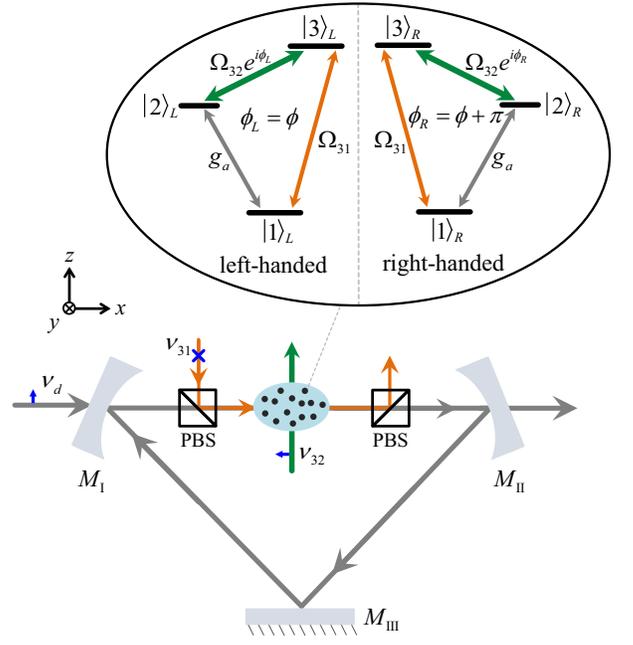}
	\caption{The model of the CQED system for cyclic three-level chiral molecules under consideration. The cavity is composed of three mirrors $M_{\rm{I}}$, $M_{\rm{II}}$, and $M_{\rm{III}}$. Here, the mirrors $M_{\rm{I}}$ and $M_{\rm{II}}$ are partially reflective and have the same reflection coefficients, while the mirror $M_{\rm{III}}$ is assumed to be perfectly reflective. The chiral molecules are coupled with the quantized cavity field and two classical light fields to form the cyclic three-level systems. The polarization directions of the light fields are shown in blue. The classical light field with angular frequency $\nu_{31}$ is introduced through the polarizing beam splitter (PBS) to interact with the molecules.}
	\label{Model}
\end{figure}

We consider the CQED system for cyclic three-level chiral molecules as shown in Fig.~\ref{Model}. The system consists of a driven traveling-wave cavity and a ensemble of chiral mixture confined in the cavity. The drive field with amplitude $\varepsilon_d$ and angular frequency $\nu_{d}$ enters the cavity from mirror $M_{\rm{I}}$ and exits from mirror $M_{\rm{II}}$. The chiral mixture contains $N=N_L+N_R$ molecules with $N_L$ and $N_R$ denoting the numbers of left- and right- handed molecules, respectively. The subscript $Q~(=L,R)$ is introduced to represent the molecular chirality. Each molecule in the chiral mixture is modeled as the cyclic three-level system, where the ground state $\left|1\right\rangle_{Q}$ is coupled to the state $\left|2\right\rangle_{Q}$ by the quantized cavity field with angular frequency $\omega_{a}$ and the state $\left|1\right\rangle_{Q}$ ($\left|2\right\rangle_{Q}$) is coupled to the state $\left|3\right\rangle_{Q}$ by the classical light field with angular frequency $\nu_{31}$ ($\nu_{32}$). Here, we focus on the three-photon resonant condition
\begin{align}
\nu_{31}=\nu_{d}+\nu_{32}.
\label{3photon}
\end{align}

Under the dipole approximation and rotating-wave approximation, the Hamiltonian in the interaction picture with respect to ${H_0} = {\nu _d}{a^\dag }a + {\nu_d}(S_{22}^L + S_{22}^R) + {\nu _{31}}(S_{33}^L + S_{33}^R)$ is written in the time-independent form as ($\hbar=1$)
\begin{align}
{H_I} =& \,{\Delta_a}{a^\dag }a + i\sqrt {{\kappa _a}} ({\varepsilon_d}{a^\dag} - \varepsilon_d^*a) \nonumber\\
&+ {\Delta_{21}}(S_{22}^L + S_{22}^R) + {\Delta_{31}}(S_{33}^L + S_{33}^R)\nonumber\\
&+[{g_a}a(S_{21}^L + S_{21}^R) + {\Omega_{31}}(S_{31}^L + S_{31}^R)  \nonumber\\
&+ {\Omega_{32}}({e^{i{\phi _L}}}S_{32}^L + {e^{i{\phi _R}}}S_{32}^R) + {\rm{H.c.}}],
\label{HamiltonianA}
\end{align}
where $\Delta_a=\omega_{a}-\nu_{d}$, $\Delta_{21}=\omega_{21}-\nu_{d}$, and $\Delta_{31}=\omega_{31}-\nu_{31}$ are the detunings, with $\omega_{21}$ and $\omega_{31}$ denoting the transition angular frequencies. $a$ ($a^\dag$) is the annihilation (creation) operator of the quantized cavity field. Here, both the cavity decay rates from mirror $M_{\rm{I}}$ and mirror $M_{\rm{I}}$ are assumed to be $\kappa_{a}/{2}$, while other cavity decay rates have been neglected. Thus, the total cavity decay rate is equal to $\kappa_{a}$. For simplicity but without loss of generality, the amplitude of the drive field, $\varepsilon_{d}$, is taken as real. $S_{jk}^{Q}=\sum_{m=1}^{N_{Q}} |j\rangle_{m\,m}^{Q\,Q} \hspace{-0.1em} \langle k|$ ($j,k=1,2,3$) are introduced to denote the collective operators for the chiral molecules. $g_a$ represents the coupling strength between the quantized cavity field and single molecules, $\Omega_{31}$ and $\Omega_{32}e^{i{\phi_Q}}$ denote the coupling strengths between the two classical light fields and single molecules. Here, $g_a$, $\Omega_{31}$, and $\Omega_{32}$ are assumed to be identical for all the molecules and are taken as real. $\phi_{L}$ and $\phi_{R}$ are the overall phases of the three coupling strengths in the cyclic three-level systems of the left- and right- handed molecules, and the chirality of the cyclic three-level system is specified as
\begin{align}
\phi_{L}=\phi,\,\phi_{R}=\phi+\pi.
\label{ChiralPhase}
\end{align}

%%%%%%%%%%%%%%%%%%%%%%%%%%%%%%%
%%%%%%%%%%%%%%%%%%%%%%%%%%%%%%%

In our CQED system, when the chiral molecules confined in the cavity are coupled with the two classical light fields, the cavity-assisted three-photon process can result in the generation of intracavity photons~\cite{CQEDMoleculeWe}. Meanwhile, the drive field can also lead to the process of the generation of intracavity photons. The interference between the intracavity photons resulting from these two processes determines the output field of the cavity, which provides a way to detect the enantiomeric excess by monitoring the output field (e.g. the transmission rate of the drive field). Therefore, our method is different from that in Ref.~\cite{CQEDMoleculeWe}, wherein only the intracavity photons generated from the cavity-assisted three-photon process determine the output field of the cavity due to the absence of external drive.

Furthermore, these collective operators $S_{jk}^{Q}$ can be expressed by introducing the generalized Holstein-Primakoff transformation~\cite{HPT-Brandes-PRL2003,HPT-Pioneer1940,HPT-RMP1991,HPT-Sun-PRL2003}~as
\begin{gather}
S_{11}^Q=N_{Q}-A_{Q}^{\dagger} A_{Q}-B_{Q}^{\dagger} B_{Q}, \ \
S_{22}^Q=A_{Q}^{\dagger} A_{Q},\ \
S_{33}^Q=B_{Q}^{\dagger} B_{Q}, \nonumber\\
S_{21}^{Q}=A_{Q}^{\dagger} \sqrt{S_{11}^Q},\ \
S_{31}^{Q}=B_{Q}^{\dagger} \sqrt{S_{11}^Q},\ \
S_{32}^{Q}=B_{Q}^{\dagger} A_{Q},
\end{gather}
where $A_{Q}$ ($A_{Q}^{\dagger}$) and $B_{Q}$ ($B_{Q}^{\dagger}$) obey the standard bosonic commutation relations $[A_{Q}, A_{Q}^{\dagger}]=[B_{Q}, B_{Q}^{\dagger}]=1$ and $[A_{Q}, B_{Q}]=[A_{Q}, B_{Q}^{\dagger}]=0$. In the low-excitation limit of molecules with large $N_Q$ limit (i.e., most molecules stay at their ground states: $\langle{A_{Q}^{\dagger}A_{Q}+B_{Q}^{\dagger} B_{Q}}\rangle\ll N_{Q}$)~\cite{HPT-3level-2003,ThreeLevel-LiY,HPT-3level-2010,HPT-3level-2013}, Hamiltonian~(\ref{HamiltonianA})~is rewritten as
\begin{align}
H_{I} \simeq &\,\Delta_a a^{\dagger}a +i\sqrt{\kappa_{a}}\varepsilon_d(a^{\dagger}-a)\nonumber\\
&+\Delta_{21} (A_{L}^{\dagger}A_{L} + A_{R}^{\dagger}A_{R}) +{\Delta _{31}}(B_{L}^{\dagger}B_{L} + B_{R}^{\dagger}B_{R}) \nonumber\\
&+[g_a a (\sqrt{N_{L}} A_{L}^{\dagger} + \sqrt{N_{R}} A_{R}^{\dagger}) + \Omega_{31}(\sqrt{N_{L}} B_{L}^{\dagger} + \sqrt{N_{R}} B_{R}^{\dagger}) \nonumber\\
&+\Omega_{32}(B_{L}^{\dagger}A_{L}e^{i \phi_{L}} + B_{R}^{\dagger}A_{R}e^{i \phi_{R}})+\mathrm{H.c.}].
\label{HamiltonianBOSEappro1}
\end{align}

In the following discussions, we will use 1,2-propanediol as an example to demonstrate our method. The working states of the cyclic three-level system are chosen as $|1\rangle=|g\rangle\left|0_{000}\right\rangle$, $|2\rangle=|e\rangle\left|1_{110}\right\rangle$, and $|3\rangle=(|e\rangle\left|1_{101}\right\rangle+|e\rangle\left|1_{10-1}\right\rangle)/\sqrt{2}$, with $|g\rangle$ ($|e\rangle$) denoting the vibrational ground (first-excited) state for the motion of OH-stretch with the transition angular frequency $\omega_{\text{vib}}=2\pi\times100.950\,\text{THz}$~\cite{PropanediolParameter1994}. The rotational states are marked in the $\left|J_{K_{a}K_{c}M}\right\rangle$ notation~\cite{Seperation-Koch-JCP,anglemomentum}. Correspondingly, as shown in Fig.~\ref{Model}, the state $\left|1\right\rangle$ is coupled to the state $\left|2\right\rangle$ by the $z$-polarized quantized field in the cavity, which is driven by the $z$-polarized classical light field. The state $\left|1\right\rangle$ ($\left|2\right\rangle$) is coupled to the state $\left|3\right\rangle$ by the $y$-polarized ($x$-polarized) classical light field. According to the rotational constants for 1,2-propanediol $A=2\pi\times8524.405\,\text{MHz}$, $B=2\pi\times3635.492\,\text{MHz}$, and $C=2\pi\times2788.699\,\text{MHz}$~\cite{PropanediolParameter2017}, the bare transition angular frequencies are obtained as $\omega_{21}=2\pi\times100.961\,\text{THz}$, $\omega_{31}=2\pi\times100.962\,\text{THz}$, and $\omega_{32}=2\pi\times0.847\,\text{GHz}$~\cite{anglemomentum}. We would like to remark that our model and method are applicable for general (asymmetric-top) gaseous chiral molecules though we take 1,2-propanediol as an example in the discussions.

\section{Steady-state transmission}\label{transmissivity}

In this section, we study the transmission of the drive field in the steady state and explore its potential applications in enantio-detection of chiral mixture.

According to Hamiltonian~(\ref{HamiltonianBOSEappro1}), one can obtain the quantum Langevin equations for the system as
\begin{align}
\dot{a}=&-K_a a-i g_{a}(\sqrt{N_{L}} A_{L}+\sqrt{N_{R}} A_{R}) +\sqrt{\kappa_{a}}(\varepsilon_{d}+a_{\rm{in}}^{\rm{I}}+a_{\rm{in}}^{\rm{II}}), \nonumber\\
\dot{A}_{Q}=&-K_A A_{Q}-i g_{a} \sqrt{N_{Q}} a -i \Omega_{32} e^{-i \phi_{Q}} B_{Q}+ F_{A}^{Q}, \nonumber\\
\dot{B}_{Q}=&-K_B B_{Q}-i \Omega_{31} \sqrt{N_{Q}} -i \Omega_{32} e^{i \phi_{Q}} A_{Q}+ F_{B}^{Q},
\label{Langevin}
\end{align}
where $K_a=i\Delta_a+\kappa_a$, $K_A=i \Delta_{21}+\Gamma_{A}$, and $K_B=i{\Delta _{31}} + \Gamma_{B}$. $a_{\rm{in}}^{\rm{I}}$ ($a_{\rm{in}}^{\rm{II}}$) is the quantum input noise operator from mirror $M_{\rm{I}}$ ($M_{\rm{II}}$) of the cavity, and has zero-mean value (i.e., $\langle a_{\rm{in}}^{\rm{I}} \rangle=\langle a_{\rm{in}}^{\rm{II}} \rangle =0$). $\Gamma_{A}$ ($\Gamma_{B}$) is introduced to denote the decay rate of the collective mode $A_Q$ ($B_Q$). $F_{A}^{Q}$ ($F_{B}^{Q}$) is the quantum input noise term of the collective operator $A_Q$ ($B_Q$), and has zero-mean value (i.e., $\langle F_{A}^{Q} \rangle=\langle F_{B}^{Q} \rangle=0$). Therefore, we obtain the following steady-state equations
\begin{align}
0=&-K_a \langle a \rangle -i g_{a}(\sqrt{N_{L}} \langle A_{L} \rangle +\sqrt{N_{R}} \langle A_{R} \rangle) +\sqrt{\kappa_{a}} \varepsilon_{d}, \nonumber\\
0=&-K_A \langle A_{Q} \rangle-i g \sqrt{N_{Q}} \langle a \rangle -i \Omega_{32} e^{-i \phi_{Q}} \langle B_{Q} \rangle, \nonumber\\
0=&-K_B \langle B_{Q} \rangle -i \Omega_{31} \sqrt{N_{Q}} -i \Omega_{32} e^{i \phi_{Q}} \langle A_{Q} \rangle,
\label{SteadyStateLangevin}
\end{align}
where $\langle O \rangle$ (with $O=a,\,A_{Q},\,B_{Q}$) represents the mean value of the operator $O$. Thus, the steady-state value of $\langle a \rangle$ is given by
\begin{align}
\langle a\rangle=&\frac{i \left(N_{L}-N_{R}\right) g_a \Omega_{31} \Omega_{32} e^{-i \phi}+\sqrt{\kappa_{a}} \varepsilon_{d}\left(K_{A} K_{B} + \Omega_{32}^{2}\right)} {K_a\left(K_{A} K_{B}+\Omega_{32}^{2}\right)+g_a^{2}N K_{B}}.
\label{aMEAN}
\end{align}
From Eq.~(\ref{aMEAN}), one can understand the physical mechanism underlying our method as follows. In the absence of the external drive (i.e., $\varepsilon_{d}=0$), only the first term in the numerator, which results from the chirality-dependent cavity-assisted three-photon process for the chiral mixture~\cite{CQEDMoleculeWe}, contributes to the intracavity photons. This term is proportional to $N_{L}-N_{R}$ since $g_a \Omega_{31} \Omega_{32} e^{-i \phi_{Q}}$ changes sign with enantiomer. When the external drive is applied (i.e., $\varepsilon_{d} \neq 0$), the second term in the numerator of Eq.~(\ref{aMEAN})~appears, resulting from the chirality-independent generation process of intracavity photons due to the drive field. Consequently, the interference between the intracavity photons generated from these two processes determines the output field of the cavity, which depends on the enantiomeric excess~\cite{Discrimination-YeC,Discrimination-XuXW,Discrimination-ChenYY}~$\eta \equiv (N_{L}-N_{R})/{N}$.

According to the input-output relation at mirrors $M_{\rm{I}}$ and $M_{\rm{II}}$ of the cavity~\cite{QuantumNoise,QuantumOpticsBook1,QuantumOpticsBook2}
\begin{align}
\sqrt{\kappa_{a}} a & = a_{\rm{in}}^{\rm{I}}+a_{\rm{out}}^{\rm{I}}+\varepsilon_{d}, \nonumber\\
\sqrt{\kappa_{a}} a & = a_{\rm{in}}^{\rm{II}}+a_{\rm{out}}^{\rm{II}},
\end{align}
one can obtain the mean output field from mirror $M_{\rm{II}}$ of the cavity $\left\langle a_{\rm{out}}^{\rm{II}}\right\rangle = \sqrt{\kappa_{a}} \left\langle a \right\rangle$. Therefore, the steady-state transmission rate of the drive field $T \equiv \left| \left\langle a_{\rm{out}}^{\rm{II}}\right\rangle / \varepsilon_{d} \right|^2$ is given by
\begin{align}
T=\frac{\kappa_{a}}{\varepsilon_{d}^2} \left|\frac{{i N g_a \Omega_{31} \Omega_{32} e^{-i \phi}} \eta + \sqrt{\kappa_{a}}\varepsilon_{d} \left(K_{A} K_{B}+\Omega_{32}^{2}\right)}{g_{a}^{2} N K_{B}+K_{a}\left(K_{A} K_{B}+\Omega_{32}^{2}\right)}  \right|^{2}.
\label{transmission}
\end{align}

In what follows, we further assume that the quantized cavity field is resonantly coupled with the transition $\left|2\right\rangle_{Q} $$\leftrightarrow$$\left|1\right\rangle_{Q}$, which means $\Delta_{a} = \Delta_{21}$. And we assume the total number of chiral molecules $N=10^8$~\cite{CQEDMolecule2,CQEDReaction1}, the decay rates of molecules $\Gamma_A/2\pi=\Gamma_B/2\pi=0.1\,\text{MHz}$~\cite{Microwave-Doyle-Nature,Microwave-Doyle-PRL}, and the total cavity decay rate~$\kappa_a/2\pi=1\,\text{MHz}$~\cite{CavityCouplingDecay-Kampschulte-2018,CavityDecay-Hoghooghi-2019}. Here, we take the weak coupling strength $\Omega_{31}/2\pi=8\,\text{kHz}$ since such a weak coupling strength usually ensures the low-excitation limit of molecules.

%%%%%%%%%%%%%%%%%%%%%%%%%%%%%%%
\begin{figure}[tbp]
	\centering
	\includegraphics[width=8.7cm]{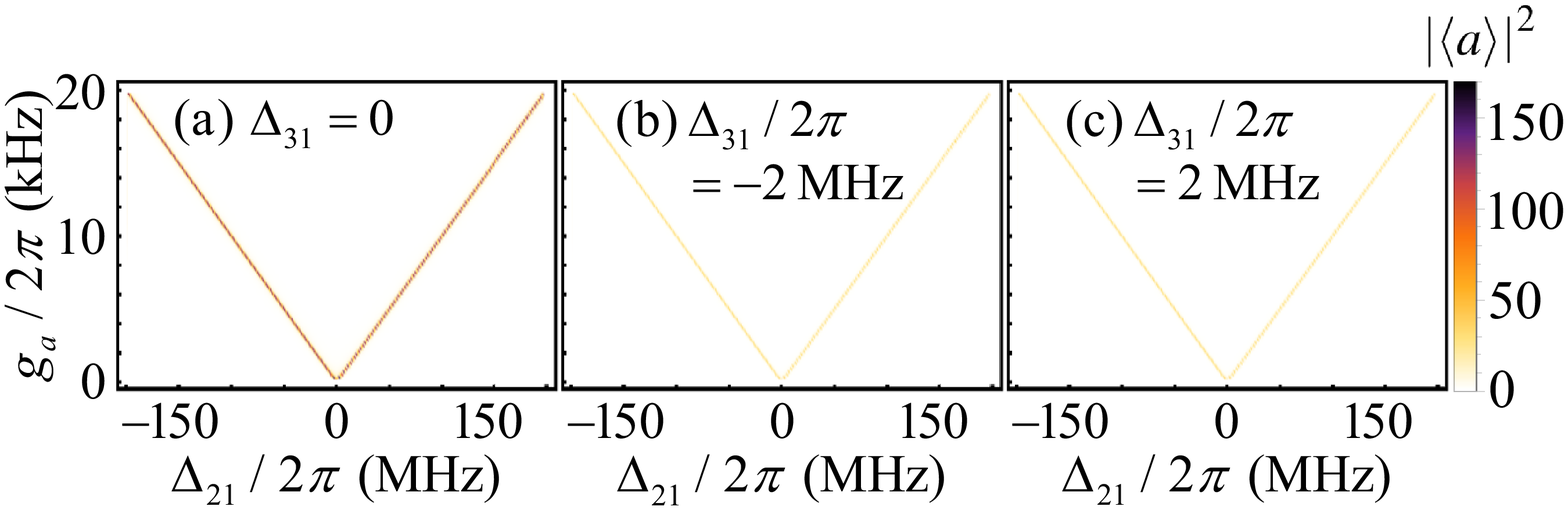}
	\caption{The steady-state intracavity mean photon number $\left| \langle a\rangle \right|^2$ in the absence of the external drive ($\epsilon_{d} = 0$) versus the detuning $\Delta_{21}$ and the coupling strength $g_a$ for (a) $\Delta_{31}=0$, (b) $\Delta_{31}/2\pi=-2\,\text{MHz}$, and (c) $\Delta_{31}/2\pi=2\,\text{MHz}$ when taking $\eta=0.9$. The other parameters are chosen as $\Delta_{a}=\Delta_{21}$, $N=10^8$, $\Gamma_A/2\pi=\Gamma_B/2\pi=0.1\,\text{MHz}$, $\kappa_a/2\pi=1\,\text{MHz}$, $\Omega_{31}/2\pi=8\,\text{kHz}$, $\Omega_{32}/2\pi=20\,\text{kHz}$, and $\phi=0$.}
	\label{Detuning}
\end{figure}
%%%%%%%%%%%%%%%%%%%%%%%%%%%%%%%

In the present work, the chirality-dependent cavity-assisted three-photon process is essential in the detection of the enantiomeric excess. Thus, we first consider the case in the absence of the external drive (i.e., $\varepsilon_{d}=0$) and display the corresponding steady-state intracavity mean photon number $\left| \langle a\rangle \right|^2$ versus the detuning $\Delta_{21}$ and the coupling strength $g_a$ for different detunings $\Delta_{31}$ in Fig.~\ref{Detuning}. As can be seen from Fig.~\ref{Detuning}(a), the intracavity mean photon number reaches the maximum $\left| \langle a\rangle \right|^2 \simeq 180$ at the detunings $\Delta_{21} \simeq \pm g_a\sqrt{N}$. This is the result of the vacuum Rabi splitting induced by the quantized cavity field in the strong collective coupling condition~\cite{QuantumOpticsBook2,RabiSplit1,CQEDtransfer2,CQEDReaction1}. Here, it is worth mentioning that since the molecules are collectively coupled to the common quantized cavity field in our system, the collective coupling strength (between the quantized cavity field and the collective mode $A_Q$) $g_a\sqrt{N}$ can be strong even the single-molecule coupling strength (between the quantized cavity field and single molecules) $g_a$ is weak. Such a collectively-enhanced coupling strength, which depends on the total number of the molecules $N$, can release the technical requirements for single-molecule strong coupling strength~\cite{CQEDtransfer2,CQEDReaction1}. Moreover, it is found that there are more intracavity photons in the resonant case $\Delta_{31}=0$ compared with the non-resonant case $\Delta_{31} \neq 0$ [see Figs.~\ref{Detuning}(a)-\ref{Detuning}(c)]. In the further discussions, we take the coupling strength $g_{a}/2\pi=10\,\text{kHz}$ and the detuning $\Delta_{31}=0$.

\begin{figure}[tbp]
	\centering
	\includegraphics[width=8.4cm]{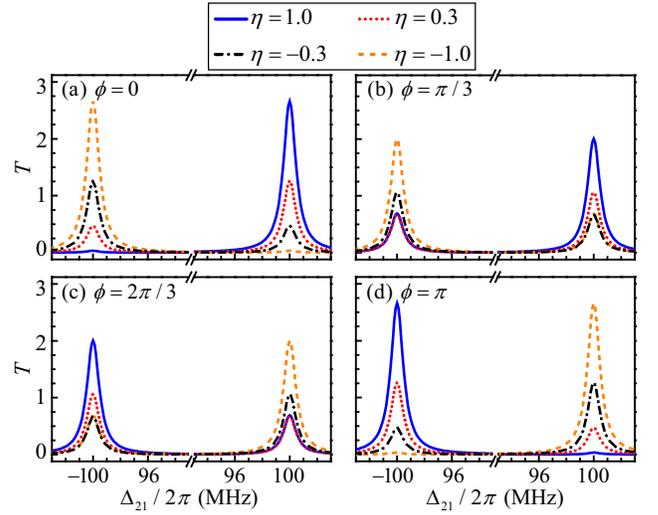}
	\caption{The transmission rate of the drive field $T$ as a function of the detuning $\Delta_{21}$ for different enantiomeric excess $\eta$ when the overall phase is taken as (a) $\phi=0$, (b) $\phi=\pi / 3$, (c) $\phi=2\pi / 3$, and (d) $\phi=\pi$. The other parameters are the same as those in Fig.~\ref{Detuning}~except $\varepsilon_{d}^2/2\pi=400\,\text{MHz}$, $g_{a}/2\pi=10\,\text{kHz}$, and ${\Delta_{31}}=0$.}
	\label{eeDepen}
\end{figure}

Furthermore, we consider the case in the presence of the external drive (i.e., $\varepsilon_{d} \neq 0$). In order to investigate the influence of the overall phase $\phi$ on the transmission rate of the drive field $T$, we choose different $\phi$ to give $T$ versus the detuning $\Delta_{21}$ in Fig.~\ref{eeDepen}. Here, only the results within the region $\phi \in[0,\,\pi]$ (e.g., $\phi=0$, $\phi=\pi / 3$, $\phi=2\pi / 3$, and $\phi=\pi$) are displayed since the results corresponding to left- and right- handed molecules will exchange when replacing the overall phase $\phi$ in the region $\phi \in[0,\,\pi]$ by $\phi+\pi$. Here, we find the transmission rate can be larger than one (i.e., $T>1$). This is the result of the constructive interference between the intracavity photons resulting from the cavity-assisted three-photon process and those generated from the drive field. Moreover, it is also shown that the transmission rate of the drive field is dependent on the overall phase. The underlying physics is that the interference between the intracavity photons generated from the cavity-assisted three-photon process and those resulting from the drive field strongly depends on the overall phase $\phi$ [see Eq.~(\ref{aMEAN})]. Specially, for the overall phase $\phi=n \pi$ (with $n$ an arbitrary integer) [see Figs.~\ref{eeDepen}(a)~and~\ref{eeDepen}(d)], $T$ becomes relatively sensitive to $\eta$ compared with the case of other values of $\phi$.

%%%%%%%%%%%%%%%%%%%%%%%%%%%%%%%
%%%%%%%%%%%%%%%%%%%%%%%%%%%%%%%

\section{Detection of enantiomeric excess}\label{DetectEE}

In this work, we focus on detecting the enantiomeric excess $\eta$ via measuring the transmission rate of the drive field $T$. On one hand, we expect that a given $T$ corresponds to only a unique $\eta$. That means $\eta$ can be detected via monitoring $T$ without requiring the enantiopure samples. On the other hand, we expect to achieve high resolution of detection, which requires that $T$ varies significantly with $\eta$.

In the following simulations of this section, the detuning $\Delta_a$, the total number of chiral molecules $N$, the decay rates of molecules ($\Gamma_A$ and $\Gamma_B$), and the coupling strength $\Omega_{31}$ are taken as the same values as those in Sec.~\ref{transmissivity}. Moreover, we assume the coupling strength $g_a/2\pi=10\,\text{kHz}$ and the detuning $\Delta_{31}=0$.

As discussed above (see Fig.~\ref{eeDepen}), at the detunings $\Delta_{21} \simeq \pm g_{a}\sqrt{N}$, the transmission rate of the drive field for $\phi=n \pi$ is relatively sensitive to the enantiomeric excess compared with the case of other values of $\Delta_{21}$. Therefore, for simplicity, we here focus on the optimal transmission rate at $\Delta_{21}=g_{a}\sqrt{N}$:
\begin{equation}
T_{\rm{op}} \simeq
\frac{\kappa_{a}}{\varepsilon_{d}^2}
\left( \frac{\sqrt{\kappa_{a}}\varepsilon_{d}\Gamma_B \pm \sqrt{N}\Omega_{31}\Omega_{32}{\eta}}{\Gamma_A\Gamma_B+\kappa_{a}\Gamma_B+\Omega_{32}^2} \right)^2,
\label{Tpeak}
\end{equation}
which is obtained by substituting $\Delta_{21}=g_{a}\sqrt{N}$ into Eq.~(\ref{transmission}) and considering the approximation $g_{a}\sqrt{N} \gg \{\kappa_{a}, \Gamma_A, \Gamma_B, \Omega_{32}, \Omega_{31} \}$. In the numerator of Eq.~(\ref{Tpeak}), ``$+$'' and ``$-$'' correspond respectively to the case of $\phi=2n\pi$ and $\phi=(2n+1)\pi$. It is found from Eq.~(\ref{Tpeak})~that, when the parameters satisfy the condition
\begin{align}
\sqrt{\kappa_{a}}\varepsilon_{d}\Gamma_B \geq \sqrt{N}\Omega_{31}\Omega_{32},
\label{Requirement1}
\end{align}
$T_{\rm{op}}$ varies with $\eta$ monotonically. That means, a given transmission rate corresponds to only a unique enantiomeric excess.

%%%%%%%%%%%%%%%%%%%%%%%%%%%%%%%
\begin{figure}[tbp]
	\centering
	\includegraphics[width=8.8cm]{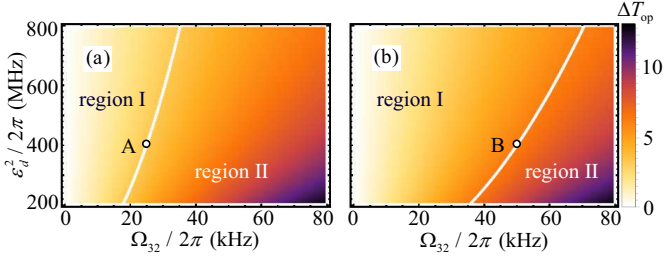}
	\caption{$\Delta T_{\rm{op}}$ in Eq.~(\ref{Requirement2})~versus the coupling strength $\Omega_{32}$ and the intensity of the drive field $\varepsilon_{d}^2$ when the total cavity decay rate is taken as (a) $\kappa_a/2\pi=1\,\text{MHz}$ and (b) $\kappa_a/2\pi=4\,\text{MHz}$. In region I where the condition $\sqrt{\kappa_{a} }\varepsilon_{d}\Gamma_B \geq \sqrt{N}\Omega_{31}\Omega_{32}$ is satisfied, $T_{\rm{op}}$ varies with $\eta$ monotonically. In region II where the condition $\sqrt{\kappa_{a} }\varepsilon_{d}\Gamma_B < \sqrt{N}\Omega_{31}\Omega_{32}$ is satisfied, $T_{\rm{op}}$ varies with $\eta$ non-monotonically. The other parameters are chosen as $\Delta_{a}=\Delta_{21}$, $N=10^8$, $\Gamma_A/2\pi=\Gamma_B/2\pi=0.1\,\text{MHz}$, $\Omega_{31}/2\pi=8\,\text{kHz}$, $g_a/2\pi=10\,\text{kHz}$, $\Delta_{31}=0$, $\Delta_{21}/2\pi=100\,\text{MHz}$, and $\phi=0$.}
	\label{Resolution}
\end{figure}
%%%%%%%%%%%%%%%%%%%%%%%%%%%%%%%

According to Eqs.~(\ref{Tpeak}) and~(\ref{Requirement1}), we further introduce the difference between the optimal transmission rates for purely left-handed ($\eta=1$) and purely right-handed ($\eta=-1$) chiral mixtures
\begin{equation}
\Delta T_{\rm{op}} = T_{\rm{op}}\lvert_{\eta=1}- T_{\rm{op}}\lvert_{\eta=-1}
\label{Requirement2}
\end{equation}
to evaluate the resolution of detection, where $T_{\rm{op}}\lvert_{\eta=1}$ ($T_{\rm{op}}\lvert_{\eta=-1}$) is obtained by substituting $\eta=1$ ($\eta=-1$) into Eq.~(\ref{Tpeak}). In Fig.~\ref{Resolution}(a), we display $\Delta T_{\rm{op}}$ versus the coupling strength $\Omega_{32}$ and the intensity of the drive field $\varepsilon_{d}^2$. Here, we take the overall phase $\phi=0$. It is shown that $\Delta T_{\rm{op}}$ strongly depends on $\Omega_{32}$ and $\varepsilon_{d}^2$. Specially, for the total cavity decay rate $\kappa_a/2\pi=1\,\text{MHz}$, one finds $\Delta T_{\rm{op}} \simeq 3$ when taking $\Omega_{32}/2\pi\simeq 25\,\text{kHz}$ and $\varepsilon_{d}^2/2\pi\simeq 400\,\text{MHz}$ [see point A in Fig.~\ref{Resolution}(a)]. For the larger total cavity decay rate $\kappa_a/2\pi=4\,\text{MHz}$, one obtains $\Delta T_{\rm{op}} \simeq 3.8$ when taking $\Omega_{32}/2\pi\simeq 50\,\text{kHz}$ and $\varepsilon_{d}^2/2\pi\simeq 400\,\text{MHz}$ [see point B in Fig.~\ref{Resolution}(b)].

Based on the results in Fig.~\ref{Resolution}, we further take $\kappa_a/2\pi=4\,\text{MHz}$, $\Omega_{32}/2\pi=50\,\text{kHz}$, and $\varepsilon_{d}^2/2\pi= 400\,\text{MHz}$ to display the optimal transmission rate $T_{\rm{op}}$ as a function of the enantiomeric excess $\eta$ for different overall phases. It is found in Fig.~\ref{PeakDenpendEE}(a)~that, for $\phi=0$ ($\phi=\pi$), $T_{\rm{op}}$ is relatively sensitive to $\eta$ in the region $\eta \in(0,\,1)$ [$\eta \in(-1,\,0)$] compared with the case in the region $\eta \in(-1,\,0)$ [$\eta \in(0,\,1)$]. Therefore, to ensure that the enantiomeric excess can be detected accurately via monitoring the transmission rate, the overall phase should be adjusted as $\phi=0$ ($\phi=\pi$) when the left-handed (right-handed) molecules are dominant in the chiral mixture.

\section{Discussions}\label{discussion}

\begin{figure}[tbp]
	\centering
	\includegraphics[width=8.6cm]{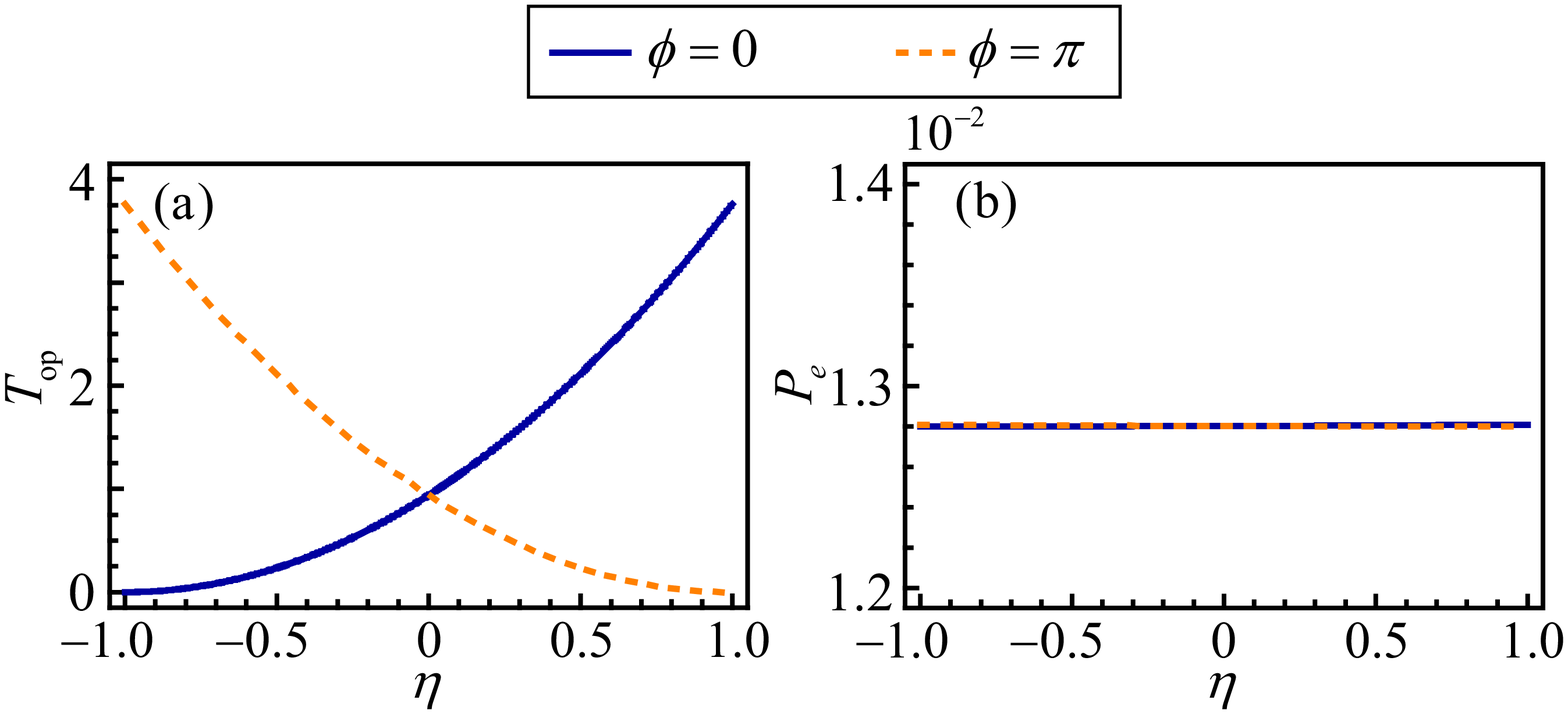}
	\caption{(a) The optimal transmission rate $T_{\rm{op}}$ and (b) the factor $P_{e}$ versus the enantiomeric excess $\eta$ for different overall phases. The other parameters are the same as those in Fig.~\ref{Resolution} except $\kappa_a/2\pi=4\,\text{MHz}$, $\Omega_{32}/2\pi=50\,\text{kHz}$, and $\varepsilon_{d}^2/2\pi=400\,\text{MHz}$.}
	\label{PeakDenpendEE}
\end{figure}

Here, it is worth mentioning that the above results are based on the low-excitation limit of molecules with large $N_Q$ limit (i.e., $\langle{A_{Q}^{\dagger}A_{Q}+B_{Q}^{\dagger} B_{Q}} \rangle \ll N_{Q}$). Thus, we introduce the factor
\begin{align}
P_e=\frac{\langle A_{L}^{\dagger} A_{L} + B_{L}^{\dagger} B_{L} \rangle}{N_{L}} + \frac{\langle A_{R}^{\dagger} A_{R} + B_{R}^{\dagger} B_{R} \rangle}{N_{R}}
\label{PE}
\end{align}
to verify whether or not the parameters used for simulations meet such a limit. The first (second) term in Eq.~(\ref{PE}) denotes the proportion of left- (right-) handed molecules occupying their excited states to the total ones $N_L$ ($N_R$). Here, we take the mean-field approximation~\cite{QuantumOpticsBook2}~$\langle A_{Q}^{\dagger}A_{Q} \rangle \simeq \langle A_{Q}^{\dagger}\rangle\langle A_{Q} \rangle$ and $\langle B_{Q}^{\dagger}B_{Q} \rangle \simeq \langle B_{Q}^{\dagger}\rangle\langle B_{Q} \rangle$. The steady-state solutions $\langle A_{Q}\rangle$, $\langle A_{Q}^{\dagger}\rangle$, $\langle B_{Q}\rangle$, and $\langle B_{Q}^{\dagger}\rangle$ are obtained by solving the steady-state Eq.~(\ref{SteadyStateLangevin}). For the parameters in Fig.~\ref{PeakDenpendEE}(a), we find $P_e \simeq 1.28 \times 10^{-2}$ [see Fig.~\ref{PeakDenpendEE}(b)]. This implies that most molecules stay at their ground states, and thus, the results meet the requirement for the low-excitation limit of molecules.

%%%%%%%%%%%%%%%%%%%%%%%%%%%%%%%
%%%%%%%%%%%%%%%%%%%%%%%%%%%%%%%

Note that in the previous CQED system for cyclic three-level chiral molecules in the standing-wave cavity~\cite{CQEDMoleculeWe}, the finite size of the sample usually would lead to the phase-mismatching problem~\cite{Discrimination-Lehmann}~and the space-dependent coupling strength~\cite{QuantumOpticsBook1}~between the quantized cavity field and single molecules. In order to evade the influence of the phase-mismatching and the spatial dependence of the coupling strength, the size of sample $l$ should be much smaller than the wavelengths of all the light fields, that is, $\{ |\vec{k}_{a}|,\,|\vec{k}_{32}|,\,|\vec{k}_{31}| \}l \ll 2\pi$. In the present method, we use the system for three-level chiral molecules confined in the traveling-wave cavity~\cite{RingCavity-Xiao-2001,RingCavity-Xiao-2008,RingCavity-Culver-2016}. When the spatial distribution of molecules is considered, only $g_a$ should be replaced by $g_a e^{i\Delta \vec{k}\cdot \vec{r}_m}$ (with the coupling strengths $\Omega_{31}$ and $\Omega_{32}$ remaining unchanged) to investigate the influence of the phase-mismatching with $\Delta \vec{k} = \vec{k}_{31}-\vec{k}_{a}-\vec{k}_{32}$, where $\vec{r}_m$ is the position of the $m$-th molecule. To ensure that the influence of the phase-mismatching is negligible, the size of sample $l$ should meet the requirement $| \Delta \vec{k} |l \ll 2\pi$. In the present system, $\vec{k}_{32}$ is the smallest one among the three vectors. Thus, with taking $\vec{k}_{31}$ and $\vec{k}_{a}$ to be parallel and $\vec{k}_{32}$ to be perpendicular to them (see Fig.~\ref{Model}), one can minimize the effect of the phase-mismatching. Here, we obtain $|\Delta \vec{k}| \simeq 2\pi \times 4.277\,\text{m}^{-1}$ for the present model of 1,2-propanediol. That means, when the sample is fixed in a volume with its size $l \ll {2\pi}/{|\Delta \vec{k}|} \simeq 0.234\,\text{m}$, the influence of the phase-mismatching can be neglected reasonably. Therefore, the requirement (i.e., the size of sample should be much smaller than the wavelengths of all the light fields) in the previous CQED method~\cite{CQEDMoleculeWe}~is released in our current method since the related cyclic three-level model is specially-designed in the traveling-wave cavity.

\section{Conclusion}\label{summary}
In conclusion, we have proposed an enantio-detection method based on the CQED system for cyclic three-level chiral molecules. The key idea is to achieve the interference between the intracavity photons generated via the cavity-assisted three-photon process and those arising from the drive field. Our results show that the enantiomeric excess can be detected via measuring the steady-state transmission rate of the drive field. In the previous CQED method~\cite{CQEDMoleculeWe}~and enantiomer-specific microwave spectroscopic methods~\cite{Microwave-Doyle-Nature,Microwave-Doyle-PRL,Microwave-Lehmann-JPCL,Microwave-Schnell-ACIE,Microwave-Schnell-JPCL}~for enantio-detection of chiral molecules, usually the enantiopure samples are required to indicate the sign of the enantiomeric excess. Note that the preparation of enantiopure samples remains a challenging work for many chiral molecules~\cite{Spatial-Separation-LiY-PRL,Spatial-Seperation-Hornberger-JCP,Spatial-Seperation-Shapiro-JCP,Spatial-Seperation-LiuB,Seperation-JiaWZ-JPB,Seperation-Koch-JCP,Seperation-LiY-PRA,Seperation-Schnell-ACIE,Seperation-SepDoyle-PRL,Seperation-Vitanov-PRR,Seperation-YeC-PRA,Seperation-ZhangQS-JPB,Conversion-Cohen-PRL,Conversion-Shapiro-JCP,Conversion-Shapiro-PRL,Conversion-YeC-PRR,Conversion-YeC1,Conversion-YeC2,Conversion-Shapiro-PRA,Seperation-Shapiro-PRL,Spatial-Seperation-Kravets-PRL,Spatial-Seperation-Cipparrone-LSAppl}. In our method, however, such enantiopure samples are not necessary since our method is based on the interference between the intracavity photons generated via the chirality-dependent cavity-assisted three-photon process and those arising from the chirality-independent drive field. Therefore, our method provides promising applications in the detection of enantiomeric excess for the chiral molecules whose enantiopure samples are still difficult to prepare.

Moreover, we note that besides enantio-detection of chiral molecules~\cite{CQEDMoleculeXia,CQEDMoleculeWe}, the CQED systems had also been used in studying energy transfer~\cite{CQEDtransfer1,CQEDtransfer2,CQEDtransfer3}~and control of chemical reactions~\cite{CQEDReaction1,CQEDReaction2,CQEDReaction3}~for molecules due to their potential applications in manipulating the molecular dynamical evolution. Therefore, in future investigations, we will further focus on the ambitious issues related to chiral molecules involving molecular dynamical evolution (such as enantio-specific state transfer, spatial enantio-separation, and enantio-conversion of chiral molecules) based on the CQED systems for cyclic three-level chiral molecules.

\begin{acknowledgments}
This work was supported by the Natural Science Foundation of China (Grants No.~12074030 and No.~U1930402), National Science Foundation for Young Scientists of China (No.~12105011), and Beijing Institute of Technology Research Fund Program for Young Scholars.
\end{acknowledgments}

\end{document}